\documentclass[twocolumn,aps,pra,floatfix,longbibliography,groupedaddress]{revtex4-1}
\usepackage{hyperref}
\usepackage{amsfonts}
\usepackage{amsbsy}
\usepackage{xcolor}
\usepackage{soul}
\usepackage{graphicx}
\usepackage{mathtools}

\newcommand{\id}{\mathbb{I}}

\newcommand{\ket}[1]{\lvert#1\rangle}

\newcommand{\abst}[1]{\lvert#1\rvert^2}

\newcommand{\Pc}{\mathcal{P}}
\newcommand{\cbE}{\boldsymbol{\mathbf{\cal E}}}

\newcommand{\unitvec}[1]{\hat{\mathbf{{#1}}}}
\renewcommand{\vec}[1]{\mathbf{#1}}

\usepackage{float}

\begin{document}
\title{Parity-time symmetry and coherent perfect absorption in a cooperative atom response}
\date{\today}
\author{K.~E.~Ballantine}
\email{k.ballantine@lancaster.ac.uk}
\author{J.~Ruostekoski}

\email{j.ruostekoski@lancaster.ac.uk}

\affiliation{Department of Physics, Lancaster University, Lancaster, LA1 4YB, United Kingdom}

\begin{abstract}

Parity-Time ($\mathcal{PT}$) symmetry has become an important concept in the design of synthetic optical materials, with exotic functionalities such as unidirectional transport and non-reciprocal reflection. At exceptional points, this symmetry is spontaneously broken, and solutions transition from those with conserved intensity to exponential growth or decay. 
Here we analyze a quantum-photonic surface formed by a single layer of atoms in an array with light mediating strong cooperative many-body interactions. We show how delocalized collective excitation eigenmodes can exhibit an effective $\mathcal{PT}$ symmetry and non-exponential decay.
This effective symmetry is achieved in a passive system without gain by balancing the scattering of a bright mode with the loss from a subradiant dark mode. These modes coalesce at exceptional points, evidenced by the emergence of coherent perfect absorption where coherent incoming light is perfectly absorbed and scattered only incoherently.
We also show how $\mathcal{PT}$ symmetry can be generated in total reflection and by balancing scattering and loss between different polarizations of collective modes.
\end{abstract}

\maketitle

\section{Introduction}

The search for novel and powerful methods to control light using artificially engineered optical properties is a current driving force in photonics~\cite{Zheludev12}. The possibility of exploiting gain and loss in photonic systems has led to interest in combined time and parity ($\mathcal{PT}$) symmetries~\cite{ElGanainy2007,Ozdemir2019,Ashida20}.
Applications of $\mathcal{PT}$ symmetric systems include, e.g., non-reciprocal light propagation~\cite{Ramezani2010,Peng2014}, unidirectional invisibility~\cite{Lin2011,Feng2013}, 
and coherent perfect absorption (CPA)~\cite{Longhi2010,Wan2011,Chong2010}. They stem from a real eigenvalue spectrum~\cite{Bender98} that undergoes spontaneous $\mathcal{PT}$ symmetry breaking at exceptional points (EPs) where eigenmodes coalesce and eigenvalues become complex~\cite{Miri2019}. Precise balance of gain and loss is technically challenging and effective $\mathcal{PT}$ symmetries in passive systems are of increasing interest~\cite{Guo09,Ornigotti14,Ozdemir2019}. In particular, incident waves can be exploited as an effective gain such that the balance between scattering and absorption loss creates $\mathcal{PT}$ symmetry~\cite{Kang2013,Sun2014}.

Metasurfaces~\cite{Yu14} are thin nanostructured films of typically subwavelength-sized scatterers for the manipulation, detection, and control of light. Placing cold atoms in a two-dimensional (2D) planar array, e.g., by an optical lattice with unit occupancy has been proposed as a nanophotonic atom analogy to a metasurface~\cite{Jenkins2012a}, consisting of essentially point-like quantum scatterers. Transmission resonance narrowing due to giant Dicke subradiance below the fundamental quantum limit of a single-atom linewidth was recently experimentally observed~\cite{Rui2020} for a planar atom array, where all the atomic dipoles oscillated in phase, while incident fields can drive giant subradiance also in resonator metasurfaces~\cite{Jenkins17}.
These quantum-photonic~\cite{Solntsev2020} surfaces of atoms that have no dissipative losses due to absorption have several advantages over artificial resonator metasurfaces (even over low-dissipation all-dielectric ones~\cite{Solntsev2020}),
as they could naturally operate in a single-photon limit~\cite{Guimond2019,Ballantine20ant,Ritsch_subr,Orioli19,Williamson2020b,Cidrim2020,Bekenstein2020,Ballantine20Huygens,Alaee20}, every atom has well-defined resonance parameters, and non-linear response~\cite{Bettles20,Parmee2020} is easily achievable. 

Here we show how the many-body dynamics of strongly coupled atoms  in a planar array, with the dipole-dipole interactions mediated by resonant light, can be engineered to exhibit CPA, associated with an effective $\mathcal{PT}$ symmetry breaking. By controlling atomic level shifts (e.g., by ac Stark shifts due to lasers), a collective analogue of electromagnetically-induced transparency (EIT) can be constructed~\cite{Facchinetti16} by coupling light dominantly only to two uniform collective excitation modes: a  `bright mode', with atomic dipoles coherently oscillating in phase in the plane of the array, and a  `dark mode', with the dipoles oscillating in phase perpendicular to the plane. As in a standard independent-atom EIT~\cite{FleischhauerEtAlRMP2005}, the bright and dark modes couple strongly and weakly to radiation, respectively. However, here the microscopic origin of the dark mode is not single-atom interference, but collective many-body subradiance.
We show that by balancing the loss of the dark mode with the scattering of the bright mode, the system becomes invariant to ${\cal PT}$ reversal.
The collective eigenmodes of the system, which are linear combinations of the in-plane and out-of-plane modes, can then coalesce at the EPs, leading to spontaneous breaking of the $\mathcal{PT}$ symmetry of the eigenvectors, realizing CPA, where coherent incoming light is perfectly absorbed and scattered only incoherently. The process can be qualitatively described by a simple two-mode model where only the two dominant collective modes are included.
We also demonstrate $\mathcal{PT}$ symmetry in other effective models of the system, related to total reflection as well as balanced scattering and loss between different polarizations of collective modes.

\section{Atom Light Coupling}

\subsection{Basic relations}\label{sec:relations}

Our model consists of a square array of cold atoms, in the $yz$ plane, with one atom per lattice site, and lattice constant $d$. The dipole moment of atom $j$ is $\vec{d}_j = \mathcal{D} \sum_\sigma \Pc_\sigma^{(j)} \unitvec{e}_\sigma$, where $\mathcal{D}$ is the reduced dipole matrix element, and $\Pc_{\sigma}^{(j)}$ and $\unitvec{e}_\sigma$ are the polarization amplitude and the unit vector associated with the assumed $\ket{J=0}\rightarrow\ket{J^\prime=1,m=\sigma}$ transition, respectively.\footnote{Here the quickly varying terms $\exp{(i\omega t)}$ have been filtered out such that the amplitudes refer to the slowly varying, positive frequency components.} We take the level detunings of this transition to be controllable, as could be achieved by ac Stark shifts of lasers or microwaves~\cite{gerbier_pra_2006}, or by magnetic fields. In the limit of low light intensity the polarization amplitudes obey~\cite{Lee16}
\begin{equation}\label{eq:Peoms}
\frac{d}{dt} \Pc_{\sigma}^{(j)}  
  =  \left( i \Delta_\sigma - \gamma \right)
  \Pc^{(j)}_\sigma + i\frac{\xi}{\mathcal{D}}\hat{\vec{e}}_\sigma^{\ast}\cdot\epsilon_0\vec{E}_{\rm ext}(\vec{r}_j),
\end{equation}
where $\Delta_\sigma=\omega-\omega_\sigma$ is the detuning of the transition to level $\sigma$ from the laser frequency $\omega=ck$, $\gamma=\mathcal{D}^2k^3/6\pi\hbar\epsilon_0$ is the single-atom Wigner-Weisskopf linewidth, and $\xi=6\pi\gamma/k^3$. The electric field consists of the incident light $
\cbE(\vec{r})  = \mathcal{E}(y,z)\unitvec{e}_y e^{ikx}$ and the sum of the scattered light from all other atoms, 
$\vec{E}_{\rm ext}(\vec{r}_j) = \cbE(\vec{r}_j) + \vec{E}_s(\vec{r}_j)$, with
\begin{equation}
\epsilon_0\vec{E}_s(\vec{r}_j) = \sum_{k\neq j} \mathsf{G}(\vec{r}_j-\vec{r}_k) \vec{d}_k.
\end{equation}
Here $\mathsf{G}$ is the standard dipole radiation kernel such that the scattered field at a point $\vec{r}$ from a dipole $\vec{d}$ at the origin is given by (for $r=|\vec{r}|$, $\unitvec{r}=\vec{r}/r$)
\begin{multline}\label{eq:rad_kernel_def}
\mathsf{G}(\vec{r})\vec{d} = \frac{\vec{d}\delta(\vec{r})}{3}+\frac{k^3 e^{ikr}}{4\pi}\left\{(\unitvec{r}\times\unitvec{d})\times\unitvec{r}\frac{1}{kr}
\right. \\
\left. -\left[ 3\unitvec{r}(\unitvec{r}\cdot\unitvec{d})-\unitvec{d} \right]\left[ \frac{i}{(kr)^2}-\frac{1}{(kr)^3} \right]
 \right\}.
\end{multline}

Inserting $\vec{E}_{\rm ext}$ into Equation~(\ref{eq:Peoms}) leads to a set of coupled equations for the polarization amplitudes~\cite{Lee16},
\begin{eqnarray}
\frac{d}{dt} \Pc_{\sigma}^{(j)}  
  =  \left( i \Delta_\sigma^{(j)} - \gamma \right)
  \Pc^{(j)}_\sigma + i\xi\sum_{k\neq j}\mathcal{G}^{(jk)}_{\sigma\tau}\Pc^{(k)}_\tau \nonumber \\ \label{eq:poleoms}
+  i\frac{\xi}{{\cal D}} \hat{\vec{e}}_{\sigma}^{\ast} \cdot
  \epsilon_0 \cbE(\vec{r}_j),
\end{eqnarray} 
where $\mathcal{G}^{(jk)}_{\sigma\tau}=\unitvec{e}_\sigma^\ast\cdot \mathsf{G}(\vec{r}_j-\vec{r}_k)\unitvec{e}_\tau$.
These can be cast in matrix form by writing the polarization amplitudes as $b_{3\sigma+j-1}=\Pc_\sigma^{(j)}$, and the drive as $f_{3\sigma+j-1}=i\xi\epsilon_0\unitvec{e}^*_\sigma\cdot\cbE(\vec{r}_j)/\mathcal{D}$ to give
\begin{equation}\label{eq:beoms}
\dot{\vec{b}} = i\mathcal{H}\vec{b}+\vec{f},
\end{equation}
where the diagonal elements $\mathcal{H}_{3\sigma+j-1,3\sigma+j-1}=(\Delta_\sigma+i\gamma)$, and the off-diagonal elements 
$\mathcal{H}_{3j+\sigma-1,3k+\tau-1} = \xi\mathcal{G}^{(jk)}_{\sigma\tau} 
$ for $j\neq k$ represent dipole-dipole interactions between different atoms.

\subsection{Collective excitation eigenmodes}

The dynamics of the atomic ensemble can be understood in terms of the collective excitation eigenmodes $\vec{v}_n$ of $\mathcal{H}$. The eigenvalues $\lambda_n=\delta_n+i\upsilon_n$ give the  collective resonance line shift $\delta_n$ and linewidth $\upsilon_n$~\cite{Jenkins_long16,Sutherland1D,Asenjo-Garcia2017a,Zhang2018,Needham19}, with $\upsilon_n<\gamma$ ($\upsilon_n>\gamma$) defining subradiant (superradiant) modes. 

The eigenvectors of the non-Hermitian $\mathcal{H}$ are not orthogonal in general, but satisfy a biorthogonal relationship between the right eigenvectors $\vec{v}_n$, $\mathcal{H}\vec{v}_n=\lambda_n\vec{v}_n$, and the left eigenvectors, defined by $\vec{w}_m^\dagger \mathcal{H}=\lambda_m\vec{w}_m^\dagger$. If $\mathcal{H}$ is diagonalizable, then we can write~\cite{Ashida20} 
\begin{equation}
\mathcal{H} = \sum_n \lambda_n \vec{v}_n\vec{w}_n^\dagger.
\end{equation}
The eigenvectors obey $\vec{w}_n^\dagger\vec{v}_m=\delta_{nm}$, but $\vec{v}_n^\dagger\vec{v}_m\neq\delta_{nm}$ in general. 
The mean Petermann factor~\cite{Petermann1979,Ashida20},
\begin{equation}\label{eq:petermanndef}
K = \frac{1}{M}\sum_{m=1}^M (\vec{w}_m^\dagger\vec{w}_m)(\vec{v}_m^\dagger\vec{v}_m),
\end{equation} 
is an important single measure of the non-orthogonality of an $M\times M$ matrix, with $K=1$ for Hermitian matrices. The overlap between right eigenvectors is bounded by~\cite{Ashida20}
\begin{equation}
|\vec{v}_n^\dagger\vec{v}_m|^2 \leq \frac{4\gamma_n\gamma_m}{(\delta_n-\delta_m)^2+(\gamma_n+\gamma_m)^2},
\end{equation}
and so at degenerate points $\lambda_n=\lambda_m$ it is possible to have an EP where two or more eigenvectors coalesce completely and become linearly dependent. The overlap between left eigenvectors appearing in Equation~(\ref{eq:petermanndef}) is not bounded, and at these points $K$ diverges. 

At such points the right eigenvectors no longer form a basis, and $\mathcal{H}$ cannot be diagonalized. It can, however, be expressed in Jordan normal form with eigenvalues on the diagonal, and ones or zeros on the superdiagonal. The solution of a homogeneous undriven ($\vec{f}=0$) set of linear equations~(\ref{eq:beoms}) is
$
\vec{b}(t) = \exp{(i\mathcal{H} t)}\vec{b}(0).
$
Only when the matrix $\mathcal{H}$ can be diagonalized the dynamics separates into independent exponential evolution of each mode. The explicit form of $\exp{(i\mathcal{H} t)}$ (Appendix) contains terms which are linear or higher order in time, depending on the degree of the degeneracy, in addition to an overall exponential dependence. 

In the absence of Zeeman splitting the atomic level structure is isotropic, and $\mathcal{H}$, while non-Hermitian, is symmetric. In this case the left and right eigenmodes are related by $\vec{w}_j^\dagger=\vec{v}^T_j$, and so $\vec{v}_j^T\vec{v}_k=0$ for $j\neq k$. While modes with $\vec{v}_j^T\vec{v}_j=0$ are possible, we can always choose a basis such that $\vec{v}_j^T\vec{v}_k=\delta_{jk}$ and can then define an occupation measure of the $j^\mathrm{th}$ mode~\cite{Facchinetti16},
\begin{equation}\label{eq:occupation}
L_j = \frac{|\vec{v}_j^T \vec{b}|^2}{\sum_k |\vec{v}_k^T\vec{b}|^2}.
\end{equation}
We find in Section~\ref{sec:Zeeman} that, while $\vec{v}_j$ are no longer eigenmodes of the full system when Zeeman splitting is turned on, it is still useful to describe the physics in this basis of eigenmodes of the unperturbed system. 

\begin{figure}[htbp]
  \centering
   \includegraphics[width=\columnwidth]{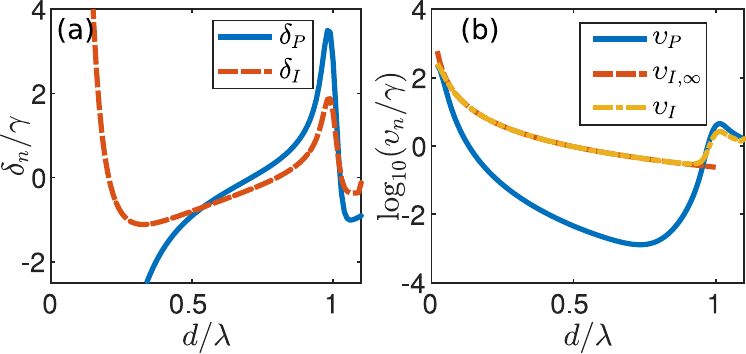}
   \vspace{-0.6cm}
  \caption{Collective excitation eigenmode (a) line shifts and (b) (log-scale) linewidths of the uniform mode with dipoles pointing out-of-plane ($\delta_P$, $\upsilon_P$) and in-plane ($\delta_I$, $\upsilon_I$) for a $20\times 20$ lattice. $\upsilon_{I,\infty}$ is the $N\rightarrow\infty$ analytic formula [Equation~(\ref{eq:inplanelinewidth})] for $\upsilon_I$. Resonance level shifts are equal for $d\approx 0.54\lambda$.  
  }
  \label{fig:delgammad}
  \end{figure}

There are two collective modes of the unperturbed system that can describe the optical response of the array~\cite{Facchinetti16}. They correspond to uniform excitations where the dipoles oscillate in phase and point either in the plane (in-plane mode $\Pc_I$), or perpendicular to the plane (out-of-plane mode $\Pc_P$). The collective line shifts, shown in Figure~\ref{fig:delgammad}(a), are generally distinct, but intersect at $d=0.54\lambda$, where $\delta_{I,P}=-0.71\gamma$. For subwavelength spacing $d<\lambda$, light can only be scattered in the single zeroth order Bragg peak in the forward or backward direction. When the dipoles point in-plane they can radiate in these directions, and so the linewidth never substantially narrows [Figure~\ref{fig:delgammad}(b)]. In the infinite-lattice limit for $d<\lambda$, we find~\cite{CAIT,Facchinetti18}
\begin{equation}
\label{eq:inplanelinewidth}
\upsilon_{I,\infty}\equiv \lim_{N\rightarrow\infty} \upsilon_I = {3\lambda^2\gamma\over 4\pi d^2},
\end{equation}
with this approximation already within $1\%$ of the numerically calculated value for $N\gtrsim 10\times 10= 100$.

The collective in-plane excitation eigenmode has now been experimentally observed~\cite{Rui2020}, with all atoms oscillating coherently in phase  in an optical lattice of $^{87}$Rb atoms with $d=532$nm and near unity filling fraction. 
According to Equation~(\ref{eq:inplanelinewidth}), this spacing with $d=0.68\lambda$ represents a subradiant collective state, resulting in a measured transmission linewidth $\approx 0.66\gamma$, well below the single-atom linewidth. 
Experiments on illuminating atoms also in optical tweezer arrays are ongoing~\cite{Glicenstein2020}.

In the out-of-plane mode the dipoles radiate predominantly sideways, away from the dipole axis, and light is scattered many times before it can escape from the edges of the lattice. This leads to a dramatically narrowed linewidth with the mode becoming completely dark for large arrays, $\lim_{N\rightarrow\infty} \upsilon_P =0$, falling off as  $\upsilon_P\propto N^{-0.9}$~\cite{Facchinetti16}.

\subsection{Zeeman shifts and two-mode model}
\label{sec:Zeeman}

The in-plane mode with a uniform phase profile can easily be excited by a wave incident in the $x$ direction normal to the plane, as also shown experimentally~\cite{Rui2020}. Deeply subradiant modes, with $\upsilon_n \ll \gamma$, are in general much harder to excite. In the case of the out-of-plane mode the dipoles point perpendicular to the plane, orthogonal to the polarization of a normally incident beam. However, $\Pc_P$ can be excited by first driving $\Pc_I$, and then transferring the excitation to $\Pc_P$ by controlling the level shifts~\cite{Facchinetti16}. To illustrate this, first consider the effect of level shifts on a single-atom. When the splitting is turned on, the isotropy of the $J=0\rightarrow J^\prime = 1$ transition is broken. The amplitudes $\Pc _{\pm}$ are driven by the circular light polarizations $\unitvec{e}_{\pm}$ with different resonance frequencies, which in the Cartesian basis appears as coupling between the atomic polarization amplitudes in the $x$ and $y$ directions. The resonances of the single-atom $x$ and $y$ component responses are shifted by an overall average detuning $\tilde{\delta} = \Delta_0-(\Delta_{+}+\Delta_{-})/2$, while the splitting between the $\sigma=\pm 1$ levels couples the Cartesian components with strength $\bar{\delta}=(\Delta_{-} - \Delta_{+})/2$, causing the dipoles to rotate. 

Similarly for the atomic lattice, the collective modes $\Pc_I$ and $\Pc_P$ of the unperturbed lattice are no longer eigenmodes of the full system when the level degeneracy is broken and are instead coupled together. The dynamics of the entire atomic response, however, can be well described by a model of only these two coupled modes, in analogy with the single-atom case~\cite{Facchinetti16}, 
\setlength{\arraycolsep}{1pt}
\begin{align}\label{eq:H0} \nonumber
\begin{pmatrix}
\dot{\Pc}_I \\ \dot{\Pc}_P
\end{pmatrix} &= i H_0 \begin{pmatrix}
\Pc_I \\ \Pc_P
\end{pmatrix}+ \begin{pmatrix}
f \\ 0
\end{pmatrix}, \\
H_0 &= \begin{pmatrix}
\tilde{\delta}_I+i\upsilon_I && -i\bar{\delta} \\ i\bar{\delta} && \tilde{\delta}_P+i\upsilon_P
\end{pmatrix},
\end{align}
where $\tilde{\delta}_{I,P}=\Delta_0+\delta_{I,P}-\tilde{\delta}$ and $f=i\epsilon_0\xi {\cal E}/\mathcal{D}$. Here the incident light directly drives only $\Pc_I$, but the level shift contribution $\bar{\delta}$ couples it indirectly to $\Pc_P$. 
\setlength{\arraycolsep}{3pt}

\begin{figure}[htbp]
  \centering
   \includegraphics[width=\columnwidth]{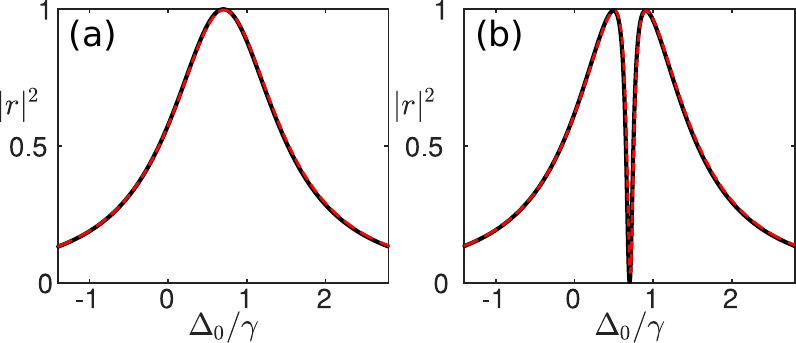}
      \vspace{-0.6cm}
  \caption{Accuracy of the two-mode model. Reflectance from a $30\times 30$ lattice calculated from two-mode model (full black line)  and full numerics (dashed red line) for (a) $\bar{\delta}=0$ where $\Pc_I$ is excited on resonance at $\Delta_0=0.71\gamma$ leading to total reflection and (b) $\bar{\delta}=0.2\gamma$ where $\Pc_P$ is excited, leading to full transmission, with $d=0.54\lambda$, $\tilde{\delta}=0$, for Gaussian input with $1/e^2$ radius $w=12 d$. Two-mode model uses parameters $\delta_{P,I}=-0.71$, $\upsilon_P=0.003$, $\upsilon_I=0.83$ extracted from numerical calculation. 
  }
  \label{fig:twomode}
  \end{figure}

The applicability of the two-mode model follows from the absence of coupling to all other collective modes due to phase mismatching. Figure~\ref{fig:twomode} shows the accuracy of the effective two-mode model in a reflection lineshape
of a Gaussian input beam from a $30\times 30$ array, exhibiting a narrow Fano resonance and variation between complete transmission and total reflection~\cite{Facchinetti16}. The reflection amplitude is given by
\begin{equation}\label{eq:ref}
r=\frac{\unitvec{e}_y\cdot \vec{E}_{S}(-\unitvec{e}_x)}{\cal E} \simeq \frac{i \upsilon_I Z_P}{\bar{\delta}^2-Z_I Z_P},
\end{equation}
where $\vec{E}_{S}(-\unitvec{e}_x)$ is the scattered field in the backward direction and the second expression is the solution of the two-mode model \eqref{eq:H0} with $Z_{I,P}=\tilde{\delta}_{I,P}+i\upsilon_{I,P}$.
As the laser frequency is tuned through the resonance $\delta_I$ of the in-plane mode for zero Zeeman shifts [Figure~\ref{fig:twomode}(a)], the atoms fully reflect the light on resonance; a well-known result for dipolar arrays~\cite{Tretyakov} and strongly influenced by collective responses~\cite{Abajo07,CAIT,Bettles2016,Facchinetti16,Shahmoon}.
Figure~\ref{fig:twomode}(b) shows the case where $\bar{\delta}^2\gg \upsilon_P\upsilon_I$, leading to excitation of $\Pc_P$ and full transmission, as this mode cannot scatter in the forward or backward direction. A sharp Fano resonance is due to the subradiant $\Pc_P$, with the collective modes reminiscent of the dark and bright states in EIT~\cite{FleischhauerEtAlRMP2005}, 
allowing light to be stored in the dark mode~\cite{Facchinetti16,Facchinetti18,Ballantine20ant}. 
The occupation measure $L$ in Equation~(\ref{eq:occupation}) for the in-plane mode is as high as $0.99$ for the Gaussian input beam considered here~\cite{Facchinetti18}.

\section{Non-exponential decay}\label{nonexp}

As a first example of the physics of EPs, we consider such points away from $\mathcal{PT}$ symmetric regions of the parameter space. We analyze the many-body dynamics of strongly coupled arrays of atoms due to light-mediated interactions that exhibits EPs, also considering the corresponding effective two-mode model.
We set $d=0.54\lambda$ such that $\delta_I=\delta_P$ and $\Delta_0=-\delta_{I,P}$, resulting in the resonance of both modes being at $\tilde{\delta}_{I,P}=-\tilde{\delta}=0$. According to the two-mode model the uniform collective eigenmodes then coalesce at the point $\bar{\delta}^2=\bar{\upsilon}^2$, where $\bar{\upsilon}=(\upsilon_I-\upsilon_P)/2$, at which point $H_0$ has only one right eigenvector. In this case the eigenvalues of $H_0$ are complex on both sides of the EP, but the EP itself can be identified by a change in the decay of the polarization amplitudes in the absence of drive. While we predominantly consider polarization dynamics in the limit of low-intensity coherent drive, it is worth noting that this non-exponential decay also applies exactly to the decay of a single-photon quantum excitation~\cite{Ballantine20ant}. 

\begin{figure}[htbp]
  \centering
   \includegraphics[width=\columnwidth]{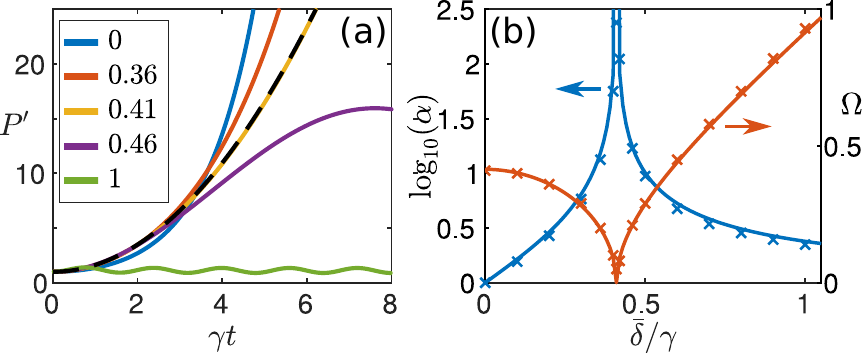}
      \vspace{-0.6cm}
  \caption{Non-exponential decay at EP of a $20\times 20$ lattice with $d=0.54\lambda$. (a) $P^\prime=e^{2\tilde{\upsilon}t}P$, with the survival probability $P=\sum|\Pc_{\sigma}^{(j)}|^2$ of a single-photon excitation from initial steady state $\Pc_I=-\Pc_P=1/\sqrt{2}$, for various $\bar{\delta}/\gamma$ given in legend. Black dashed line gives the non-exponential behavior at the EP $\bar{\delta}\approx 0.41\bar{\upsilon}$ of the corresponding two-mode model, dividing real-exponential and oscillatory, complex-exponential contributions. (b) $\alpha$ and $\Omega$ extracted from fit of numeric decay (crosses) to Equation~(\ref{eq:Pfit}), and predicted from two-mode model (lines). The EP is characterized by diverging $\alpha$ and $\Omega\rightarrow0$. }
  \label{fig:decay}
\end{figure}

We show  in Figure~\ref{fig:decay}(a) the decay of an initial state $\Pc_P=-\Pc_I$ in the absence of drive for a $20\times 20$ lattice, for varying values of $\bar{\delta}$,  with an overall exponential decay $e^{-2\tilde{\upsilon}t}$, for $\tilde{\upsilon}=(\upsilon_I+\upsilon_P)/2$, factored out. The two-mode model has an EP at $\bar{\delta}\simeq 0.41\gamma$. We find that close to this value the two collective modes of the full lattice become very similar, although the limits of numerical accuracy mean they remain distinguishable, and it also marks a transition in the remaining contribution to the decay from real-exponential growth to a complex-exponential, oscillatory form. Close to the transition point, this contribution is approximately quadratic in $t$, as illustrated by the black, dashed quadratic curve.    

The behavior as we transition through the EP is well described by the two-mode model (\ref{eq:H0}) that yields
\begin{align}\label{eq:Pfit}
P&=|\Pc_P|^2+|\Pc_I|^2\nonumber\\
&= \exp{(-2\tilde{\upsilon}t)}\left[\cos^2{(\Omega t})+\alpha\sin^2{(\Omega t)}\right] ,
\end{align}
with $\alpha=(\bar{\delta}+\bar{\upsilon})/(\bar{\delta}-\bar{\upsilon})$ and $\Omega=\sqrt{\bar{\delta}^2-\bar{\upsilon}^2}$. As $\bar{\delta}$ is varied through the EP $\bar{\delta}^2=\bar{\upsilon}^2$, $\Omega$ transitions from real to imaginary, and the term in square brackets goes between oscillatory $\sim e^{\pm i |\Omega|t}$, and exponential growth and decay $\sim e^{\pm |\Omega|t}$. Between these two regimes at the EP is non-exponential behavior, characterized by $\alpha\rightarrow\infty$, $\Omega\rightarrow 0$, with $\alpha\Omega^2\rightarrow 4\bar{\upsilon}^2$ remaining finite. Only the leading order, non-exponential contribution $~ 1+\alpha\Omega^2t^2$ in Equation~(\ref{eq:Pfit}) survives. This limit matches the expected non-exponential solution exactly at the EP;
\begin{align*}
\Pc_{-} &= \Pc_{-}(0)\exp{(-\tilde{\upsilon}t)} \\
\Pc_{+} &= \Pc_{+}(0)\exp{(-\tilde{\upsilon}t)}-2\bar{\upsilon} t \Pc_{-}(t),
\end{align*}
where $\Pc_{\pm} = (\Pc_I\pm\Pc_P)/\sqrt{2}$.

The solution of the full many-atom lattice dynamics similarly approaches the non-exponential solution at the EP, as higher-order terms in the exponential expansion are suppressed to longer and longer times (see Appendix~\ref{appdecay} for more details). Figure~\ref{fig:decay}(b) shows the parameters extracted from fitting the full numeric simulation to the form predicted by Equation~(\ref{eq:Pfit}), with $\Omega\rightarrow 0$ and $\alpha$ diverging at the EP, as expected. The mean Petermann factor calculated from the two-mode model,
\begin{equation}\label{eq:petermann}
K  =\frac{1}{2}\left(1+\frac{\bar{\delta}^2+\bar{\upsilon}^2}{|\bar{\delta}^2-\bar{\upsilon}^2|}\right),
\end{equation} 
also diverges at this point.

\section{${\cal PT}$ symmetry and CPA}
\label{sec:PTtrans}

While non-exponential decay demonstrates a physical effect of EPs even in the absence of ${\cal PT}$ symmetry, EPs are of particular interest when they coincide with the spontaneous breaking of such a symmetry, resulting in a transition of the eigenvalues from purely real to complex. In this section we show how an effective ${\cal PT}$ symmetry can arise, when scattering from the collective bright in-plane mode balances loss from the collective dark out-of-plane mode, and how this symmetry can be described by the two-mode model. At the EPs where this symmetry is spontaneously broken, solutions emerge which exhibit CPA, where all coherent transmission and reflection disappears.

\subsection{Lattice transmission and scattering}

Large arrays can respond collectively as a whole to incoming light normal to the lattice and  the atomic excitations exhibit uniform phase profiles. The amplitudes of the incoming and outgoing (transmitted and reflected) fields, as well as each of these and the mode amplitudes are then related by linear transformations in the limit of low light intensity. This leads to quasi-1D physics, where transmission through even several stacked 2D arrays can be treated as a 1D process with each lattice responding as a single `superatom' with a collective resonance line shift and linewidth~\cite{Facchinetti18,Javanainen19}. Here we show how to prepare an array in a CPA phase that corresponds to the coherently scattered field perfectly canceling the external field, leading to no outgoing coherent light. 

For zero Zeeman shifts, we can rewrite the equations of motion~(\ref{eq:H0}) in terms of the scalar amplitude of the $y$ component of the uniform mode for a single planar array~$j$
\begin{equation}\label{eq:1Deom}
\partial_t\Pc_I^{(j)} = (i\tilde\delta_I-\upsilon_I) \Pc_I^{(j)} + \frac{2i\upsilon_I}{\tilde{\mathcal{D}}k} \epsilon_0 E^{(j)}_{{\rm ext}},
\end{equation}
where $\tilde{\mathcal{D}}=\mathcal{D}/{\cal A}$ denotes the density of atomic dipoles on the array for ${\cal A}=d^2$, the element of the square unit cell, and we have used $\upsilon_I$ from Equation~(\ref{eq:inplanelinewidth}).
For a large ideal 2D subwavelength array only the in-plane mode scatters light coherently in the exact forward and backward direction. The field scattered by a planar array located at $x_j$ can then be 
approximated at $x$, where the distance from the array satisfies $\lambda \lesssim |x-x_j| \ll \sqrt{\mathcal{A}}$, by the field scattered from a uniform slab~\cite{dalibardexp,Facchinetti18,Javanainen19}. The total field is then the sum of the incident and the scattered fields
\begin{equation}
 E(\textbf{r})=\mathcal{E} e^{ i kx} + \frac{ik \tilde{{\cal D}}}{2\epsilon_0} e^{i k |x-x_j|}\Pc_I^{(j)},
\label{1dscatt}
\end{equation}
indicating that only the average dipole moment per unit area is important. This system corresponds to 1D scalar electrodynamics, and with the average  dipole density $\mathcal{D}/d^2$, the linewidth of the in-plane mode $\Pc_I$ equals the
effective 1D linewidth $\gamma_\mathrm{1D}=3\pi \gamma/(k^2{\cal A})$~\cite{Javanainen1999a}. The 3D dipole radiation kernel, Equation~(\ref{eq:rad_kernel_def}), has now been replaced by the one from 1D electrodynamics
$(ik/2) e^{i k |x-x_j|}$
which also represents the interaction between uniformly excited planar arrays at different $x$ positions. This is equivalent to  dipole-dipole coupled atoms interacting via a 1D waveguide~\cite{Facchinetti18,Javanainen19} (see Appendix~\ref{app1D}).

Here we consider the response of an individual lattice plane to a general external field and drop the label $j$. We separate the fields according to the propagation direction $E_{{\rm ext}}=E_{{\rm ext}}^{(+)}+E_{{\rm ext}}^{(-)}$, with $E_{{\rm ext}}^{(\pm)}$ denoting the amplitude of the external field coming \emph{toward} the array from the $\pm x$ direction.\footnote{All the fields refer to the positive frequency components.} The total field \emph{outgoing} from the array to the $\pm x$ direction is
\begin{equation}\label{eq:e1}
E^{(\pm)} = E_{{\rm ext}}^{(\mp)} + \eta\left(E_{{\rm ext}}^{(+)} + E_{{\rm ext}}^{(-)}\right).
\end{equation}
The second term denotes the scattered light  $E_{s}^{(\pm)}=\eta(E_{{\rm ext}}^{(+)}+E_{{\rm ext}}^{(-)})$ that is equal in both directions and depends only on the total field at the array.
The coefficient $\eta = ik\alpha/2$ can be derived by noting that, aside from the propagation phase factor, $\epsilon_0 E_{s} = (ik\tilde{\mathcal{D}}/2) \Pc_I$ [Equation~\eqref{1dscatt}], while $\tilde{{\cal D}} \Pc_I = \alpha \epsilon_0 E_{{\rm ext}} $ is determined by  the 1D polarizability $\alpha$ from the steady state of Equation~(\ref{eq:1Deom}),
\begin{equation}
\alpha = -\frac{2\upsilon_I}{k(\tilde\delta_I+i\upsilon_I)}.
\end{equation}

While an ideal lattice scatters light coherently, incoherent scattering can arise due to fluctuations in the atomic positions, quantum correlations, or defects in the lattice. 
In the following section we consider a specific numerical example of incoherent light scattering due to position fluctuations.  We now take Equation~\eqref{eq:e1} to refer only the coherent contribution to light.
Incoherently scattered light reduces the coherently scattered amplitude, the second term of Equation~\eqref{eq:e1},
and to incorporate this change, we replace $\eta$ by $\eta^\prime$ in Equation~\eqref{eq:e1}.
We then write $\eta^\prime/\eta$ as the ratio of the radiative linewidths for the coherent, denoted by $\upsilon_{I}^c$, and total scattering (coherent plus incoherent), denoted by $\upsilon_{I}^\prime$,
\begin{equation}
\eta^\prime = \frac{\upsilon_I^c \eta}{\upsilon_I^\prime},
\end{equation}
indicating that the intensity of the coherent scattering is a fraction $|\upsilon_I^c/\upsilon_I^\prime|^2$ of the total.
Consider next the steady-state versions ($\dot{\Pc}_I=\dot{\Pc}_P=0$) of Equations~(\ref{eq:H0}), with the the linewidths replaced by the primed ones $\upsilon_{I,P}^\prime$ that include 
the incoherent contributions,
\begin{equation}\label{eq:neweqn}
\begin{pmatrix}
i\upsilon_I^\prime-\tilde{\delta} && -i\bar{\delta} \\ i\bar{\delta} && i\upsilon_P^\prime-\tilde{\delta} 
\end{pmatrix}
\begin{pmatrix}
\Pc_I \\ \Pc_P
\end{pmatrix} = -\frac{2\upsilon_I'\epsilon_0}{\tilde{\mathcal{D}}k}\begin{pmatrix}
 E_{{\rm ext}} \\ 0
\end{pmatrix},
\end{equation}
where the equation for $\Pc_I$ is first transformed as in Equation \eqref{eq:1Deom}. (Again we set $\delta_I=\delta_P=-\Delta_0$ and include any additional detuning offset in $\tilde{\delta}$.) 
From Equations~(\ref{eq:e1}), together with replacing $\eta$ by $\eta^\prime$, we can substitute $ E_{{\rm ext}}=  E_{{\rm ext}}^{(+)}+  E_{{\rm ext}}^{(-)}=E^{(+)}+E^{(-)}-2\eta^\prime (E_{{\rm ext}}^{(+)}+E_{{\rm ext}}^{(-)})$ on the right-hand-side 
of Equation~\eqref{eq:neweqn}. As we are interested in the total coherent outgoing field and under which conditions this is zero, we rearrange the terms such that only this contribution appears on the right side,
\begin{equation}\label{eq:ptm}
(H_T - \tilde{\delta} \id )\begin{pmatrix}
\Pc_I \\ \Pc_P
\end{pmatrix} = -\frac{2\upsilon_I'\epsilon_0}{\tilde{\mathcal{D}}k}\begin{pmatrix}
E^{(+)} + E^{(-)} \\ 0
\end{pmatrix},
\end{equation}
where
\begin{equation}\label{eq:Ht}
H_T = \begin{pmatrix}
i\upsilon_I^\prime - 2i\upsilon_I^c && -i\bar{\delta} \\ i\bar{\delta} && i\upsilon_P^\prime 
\end{pmatrix}.
\end{equation}
The difference between this and the evolution matrix $H_0$ in Equation~(\ref{eq:H0}) is that the scattered light is expressed in terms of the mode amplitudes on the left-hand-side of Equation~\eqref{eq:ptm} (the $2i\upsilon_I^c$ term), representing passive gain due to incident light.
Parity inversion and time reversal correspond to an interchange of the two modes and complex conjugation, respectively. While $H_0$ in Equation~(\ref{eq:H0}) cannot exhibit ${\cal PT}$ symmetry for $\upsilon_{I,P}>0$, the matrix $H_T$ is invariant under this symmetry if $2\upsilon_I^c-\upsilon_I^\prime = \upsilon_P^\prime$, which is possible when $\upsilon_I^c>\upsilon_I^\prime/2$. When this symmetry is satisfied, i.e.\ when the net coherent scattering of light from the bright, in-plane mode balances the loss due to incoherent scattering from the dark, out-of-plane mode, the matrix $H_T$ is ${\cal PT}$ symmetric. The eigenvectors of $H_T$ coalesce at an EP at $\bar{\delta}^2 = (\upsilon_P^\prime)^2 = (2\upsilon_I^c-\upsilon_I^\prime)^2$. While the matrix retains the ${\cal PT}$ symmetry regardless of the value of $\bar{\delta}$, it is at this point that the eigenmodes themselves transition to being no longer invariant under the ${\cal PT}$ operation.

The scattering matrix $S$ links the total outgoing fields to the incoming fields and can also be directly derived from Equations~\eqref{eq:e1}, or alternatively from Equations~\eqref{eq:ptm}. Written in terms of the complex reflection amplitude $r$ and transmission amplitude $t$, $S$ is defined as
\begin{equation}\label{eq:smatrix}
\begin{pmatrix}
E^{(+)} \\ E^{(-)} 
\end{pmatrix} = S \begin{pmatrix}
E_{{\rm ext}}^{(-)} \\ E_{{\rm ext}}^{(+)}
\end{pmatrix}=\begin{pmatrix}
t && r \\ r && t
\end{pmatrix}\begin{pmatrix}
E_{{\rm ext}}^{(-)} \\ E_{{\rm ext}}^{(+)}
\end{pmatrix}.
\end{equation}   
It follows that
\begin{equation}\label{eq:detS}
\det{S} \propto \frac{\det{(H_T-\tilde{\delta}\id)}}{\det{H_0}},
\end{equation}
and so when $\det{(H_T-\tilde{\delta}\id)}=0$, which is possible only when $H_T$ has a real eigenvalue $\tilde{\delta}$, the scattering matrix $S$ has a zero eigenvalue. Physically, this implies there is a combination of incident light which is completely absorbed, leading to zero coherent outgoing field. At the EPs, the eigenvalues of $H_T$ transition from fully real to complex conjugate pairs, as the $\mathcal{PT}$ symmetry is spontaneously broken. When the eigenvalues are real, CPA is possible. As the scattered field in both directions is equal, full destructive interference can only occur on both sides if the incident field is also symmetric, and so the eigenvalue which goes to zero must be $t+r$, corresponding to $E_{{\rm ext}}^{(+)}=E_{{\rm ext}}^{(-)}$.

\subsection{Tuning collective mode linewidths}

As shown in the previous section, ${\cal PT}$ symmetry can be achieved when $2\upsilon_I^c-\upsilon_I^\prime=\upsilon_P^\prime$. For an ideal lattice with fixed atomic positions $\upsilon_P\ll \upsilon_I$. However, while the atoms remain in the ground state of the trap, the wavefunction in this state has a finite size. We model a finite optical lattice with depth $sE_R$, in units of the lattice photon recoil energy $E_R=\pi^2\hbar^2/(2md^2)$, by stochastically sampling the atomic positions from the harmonic oscillator Gaussian density distribution at each site, with root-mean-square width $l=ds^{-1/4}/\pi$, over many realizations~\cite{Lee16}. Then fluctuations in the atomic positions will both increase $\upsilon_P^\prime$, and tune the rate $\upsilon_I^c$ of coherent scattering. The total outgoing power, $P=P_{\rm c}+P_{\rm inc}$, has coherent contributions arising from interference between the incoming light and the average scattered field,
\begin{equation}
P_{\rm c} = 2\epsilon_0 c \int \mathrm{d}S\, \abst{\cbE + \langle \vec{E}_s\rangle},
\end{equation}
and incoherent contributions due to fluctuations,
\begin{equation}
P_{\rm inc} = 2\epsilon_0 c \int \mathrm{d}S\,\left( \langle \vec{E}^\dagger_s\cdot  \vec{E}_s\rangle-\abst{\langle \vec{E}_s\rangle}\right),
\end{equation} 
where the angled brackets represent averaging over many stochastic realizations, and the integral is over the full closed surface~\cite{Bettles20}.
As energy is radiated away, the amplitude decays with an effective scattering linewidth $\upsilon\propto P$. Incoherent scattering from both modes increases with increasing position uncertainty $l$, while coherent scattering from $\Pc_I$ decreases. This implies that while $\upsilon_{P}^\prime$ increases, the difference $2\upsilon_I^c-\upsilon_I^\prime$ decreases, and the two values approach and become approximately equal leading to an effective ${\cal PT}$ symmetry.  

\begin{figure}[htbp]
  \centering
   \includegraphics[width=\columnwidth]{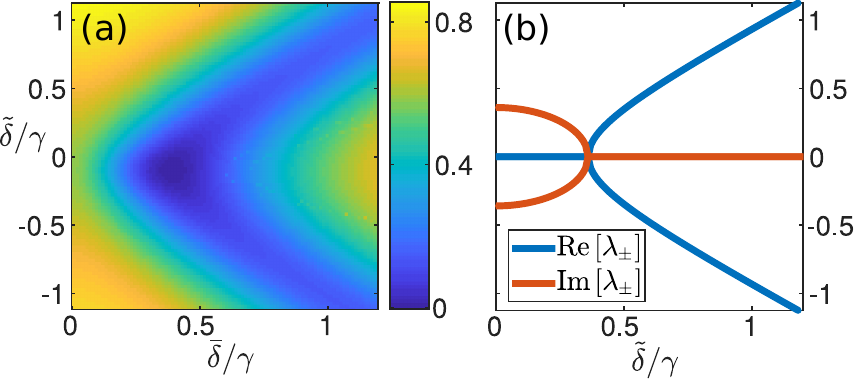}
      \vspace{-0.6cm}
  \caption{Emergence of CPA due to ${\cal PT}$ symmetry breaking. (a) Intensity $|t+r|^2$ of total coherent outgoing light for symmetric incident light showing the emergence of two real detunings corresponding to CPA. (b) Real part (blue) and imaginary part (orange) of eigenvalues of matrix $H_T$ in Equation~(\ref{eq:Ht}) (in units of $\gamma$), showing bifurcation at the EP $\bar{\delta}=(2\upsilon_I^c-\upsilon_I^\prime)=\upsilon_P^\prime=0.4\gamma$, with real eigenvalues corresponding to CPA phase shown in (a).  
  }
  \label{fig:ep1}
  \end{figure}

We calculate directly the total coherent output intensity, averaged over $1500$ stochastic realizations, for a $20\times 20$ lattice under steady-state symmetric Gaussian beam illumination from both sides, with $1/e^2$ radius $8d$, and $l=0.2d$, corresponding to $s\approx 6.5$. The output intensity, normalized by the total input intensity from both beams, gives $|t+r|^2$, shown in Figure~\ref{fig:ep1}(a). For $\bar{\delta}\approx 0.4\gamma$, approximate CPA emerges, with $|t+r|^2\approx0.03$ at the minimum. For larger $\bar{\delta}$, the CPA phase then splits into two branches. While there is considerable incoherent scattering due to position fluctuations, the total coherent output remains relatively high away from these branches.

Again this behavior is explained by the simple two-mode model. The eigenvalues $\lambda_{\pm}$ of $H_T$ are shown in Figure~\ref{fig:ep1}(b), for $\upsilon_P^\prime=2\upsilon_I^c-\upsilon_I^\prime=0.4\gamma$, corresponding to the point of maximum absorption in the numerics. At $\bar{\delta}=\upsilon_P^\prime$ there is an EP where the eigenvalues transition from purely imaginary to purely real. Above this point, the CPA phase in the full lattice is seen when $\tilde{\delta}$ matches one of these real eigenvalues, i.e.\ when $\tilde{\delta}^2=\bar{\delta}^2-(\upsilon_P^\prime)^2$. 
Equation~(\ref{eq:detS}) then implies that there are real detunings $\tilde{\delta}$ which lead to the zero eigenvalues $t+r=0$, resulting in CPA.  $K$ has the same form as Equation~(\ref{eq:petermann}), with $\bar{\upsilon}$ being replaced by $\upsilon_P^\prime$, and diverges at the EP.

\begin{figure}[htbp]
  \centering
   \includegraphics[width=\columnwidth]{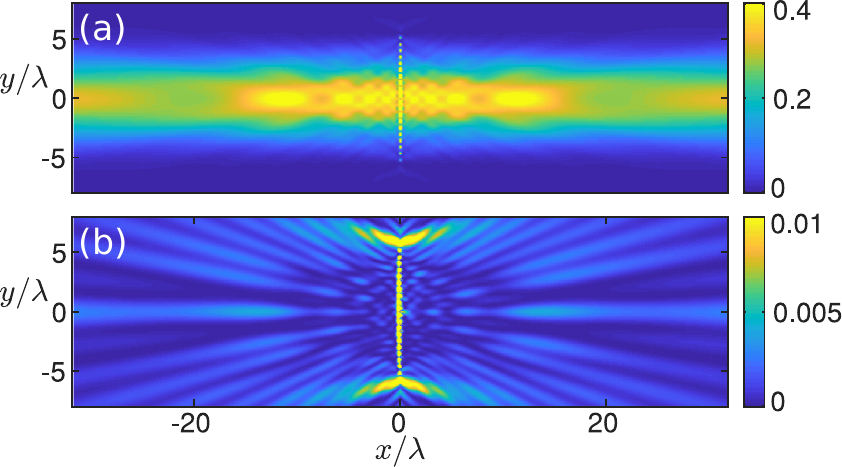}
      \vspace{-0.6cm}
  \caption{
  Spatial dependence of intensity for reflecting and CPA phases for a 20$\times$20 atom array on the $yz$ plane at $x=0$. Total outgoing coherent intensity for $\tilde{\delta}=0$, (a) $\bar{\delta}=0$, where $|t+r|^2=0.6$, and (b) $\bar{\delta}=0.4$, where $|t+r|^2=0.03$, showing CPA, when the atoms are illuminated by Gaussian beams propagating from $\pm x$ directions. Note different color scales. Field is normalized such that $|\cbE|=1$ at $x=0$. 
  }
  \label{fig:cpaspace}
  \end{figure}

The spatial variation of the total coherent outgoing field, for the same lattice parameters and illumination as Figure~\ref{fig:ep1}, is shown in Figure~\ref{fig:cpaspace} for two cases, one where the lattice is strongly reflective, and one where CPA is achieved. In the second case, there is near total cancellation between the coherent scattering and the external field, resulting in a drastic reduction in intensity, with peak intensity around the edge of the lattice.

\section{$\mathcal{PT}$ symmetry and polarization conversion}

We propose two other realizations of ${\cal PT}$  symmetric scattering with atomic arrays. The first, described here, can be achieved by considering scattering between incoming and outgoing light of a single polarization, while treating the orthogonally polarized outgoing light as loss. A final example, where full reflection from the array is described by ${\cal PT}$ symmetry, is presented in Appendix~\ref{appref}.

We now consider an array in the $xy$ plane, where the quantization axis remains aligned with the $z$ axis, such that the two in-plane uniform modes, with $\Pc_{Ix}$ and $\Pc_{Iy}$ the polarization in the $x$ and $y$ direction, respectively, are coupled. These two modes are degenerate with $\delta_{Ix}=\delta_{Iy}=\delta_I$ and $\upsilon_{Ix}=\upsilon_{Iy} = \upsilon_I$. If we consider the transmission and reflection of only $y$-polarized light, then any $x$-polarized scattering can be treated as loss. In the case of an ideal lattice, the modes $\Pc_{Ix}$ and $\Pc_{Iy}$ will each scatter light in the respective polarization, with only coherent scattering present, giving $\upsilon_{Iy}^c=\upsilon_{Ix,y}^\prime=\upsilon_I$ and $\upsilon_{Ix}^c=0$. We consider plane wave illumination $\cbE=(E_{\rm ext}^{(-)}e^{ikz}+E_{{\rm ext}}^{(+)}e^{-ikz})\unitvec{e}_y$. Due to coupling between the $y$ and $x$ directions, the scattered field will contain both polarization components, such that the total field can be decomposed into $E^{(\pm)}_{x}=E^{(\pm)}_{s,x}$ and $E^{(\pm)}_{y}=E^{(\pm)}_{s,y}+E^{(\mp)}_{{\rm ext}}$, respectively. Then the mode amplitudes for the $y$ polarization are linked to the field amplitudes, in analogy with Equation~(\ref{eq:Ht}), through
\begin{equation}\label{eq:ptmpol}
(H_{\rm pol} - \tilde{\delta} \id )\begin{pmatrix}
\Pc_{I,y} \\ \Pc_{I,x}
\end{pmatrix} = -\frac{2\upsilon_I\epsilon_0}{\tilde{\mathcal{D}}k}\begin{pmatrix}
E^{(+)}_{y} + E^{(-)}_{y} \\ 0
\end{pmatrix},
\end{equation}
where
\begin{equation}
H_{\rm pol} = \begin{pmatrix}
- i\upsilon_I && -i\bar{\delta} \\ i\bar{\delta} && i\upsilon_I 
\end{pmatrix}
\end{equation}
is automatically $\mathcal{PT}$ symmetric, with EPs at $\bar{\delta}^2=\upsilon_I^2$. 

\begin{figure}[htbp]
  \centering
   \includegraphics[width=\columnwidth]{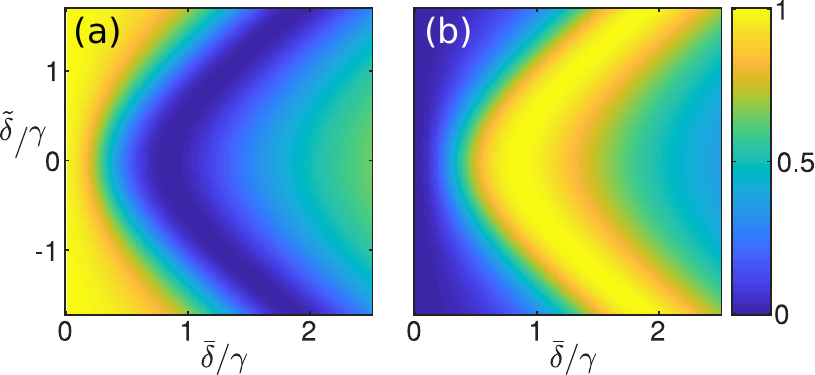}
      \vspace{-0.6cm}
  \caption{${\cal PT}$ symmetry in polarization conversion. (a) $y$ component $\abst{E^{(+)}_{y}}+\abst{E^{(-)}_{y}}$ of total outgoing (transmitted and reflected) light leaving the array as a function of $\tilde{\delta}$ and $\bar{\delta}$, and (b) outgoing $x$ polarized light $\abst{E^{(+)}_{x}}+\abst{E^{(-)}_{x}}$, for symmetric $y$ polarized plane wave incident on $20\times 20$ lattice with $d=0.54\lambda$. At the EP at $\bar{\delta}=\upsilon_I$, solutions emerge for which total polarization conversion is possible. EP occurs at $\tilde{\delta}=0$ and $\bar{\delta}=\upsilon_I=0.83\gamma$. }
  \label{fig:polconv}
  \end{figure}

Again, this $\mathcal{PT}$ symmetry leads to a zero in the scattering matrix between $y$-polarized input and output, corresponding to zero outgoing $y$ polarization for two real detunings $\tilde{\delta}$ above the EP. At these points the light is not scattered incoherently, but rather converted completely to the orthogonal $x$ polarization. Figure~\ref{fig:polconv} shows the outgoing intensity in the $y$ and $x$ polarizations for symmetric $y$-polarized incidence, with perfect coherent polarization conversion emerging above the EP.

\section{Conclusion}

${\cal PT}$ symmetry has emerged as an important topic in the physics of non-Hermitian systems, as this symmetry can lead to real eigenvalues of the system Hamiltonian. The behavior of a ${\cal PT}$ symmetric system can change dramatically at exceptional points, where eigenvectors coalesce and this symmetry is spontaneously broken. This is in stark contrast to Hermitian systems, where eigenvectors remain orthogonal even at spectral degeneracies.

We have shown how an effective ${\cal PT}$ symmetry can arise, without gain, in the many-body dynamics of strongly coupled arrays of atoms due to light-mediated interactions, leading to CPA with almost no coherent transmission or reflection. This symmetry is achieved by balancing the coherent scattering from a bright mode, which couples strongly to incident radiation, with the incoherent scattering from a dark mode, which couples weakly. The ${\cal PT}$ symmetry is spontaneously broken at EPs where the eigenmodes coalesce, and the CPA phase emerges, with non-exponential decay at these points. Remarkably, the optical response of arrays of hundreds of atoms can be described by a simple two-mode model, which explains the complete coherent absorption, reflection, and polarization conversion of light, providing a simple and intuitive description. Here, however, the modes in question are collective modes of the entire many-body system, whose properties arise from cooperative interactions and differ dramatically from that of a single atom.

\begin{acknowledgments}
We acknowledge financial support from the UK EPSRC (Grant Nos. EP/S002952/1, EP/P026133/1)
\end{acknowledgments}

\appendix

\section{Non-exponential decay and exceptional points}\label{appdecay}

We analyze the decay near the exceptional point of Sec.~\ref{nonexp} in the absence of drive, without ${\cal PT}$ symmetry. Assuming $\Delta_0+\delta_{I,P}-\tilde{\delta}=0$, the equations of motion of the two-mode model are
\begin{align}
\partial_t\begin{pmatrix}
\Pc_I \\ \Pc_P
\end{pmatrix} &= \begin{pmatrix}
- \upsilon_I && \bar{\delta} \\ -\bar{\delta} && -\upsilon_P
\end{pmatrix} \begin{pmatrix}
\Pc_I \\ \Pc_P
\end{pmatrix} \\
&= \left[ -\tilde{\upsilon}\id +  \begin{pmatrix}
- \bar{\upsilon} && \bar{\delta} \\ -\bar{\delta} && \bar{\upsilon}
\end{pmatrix}\right] \begin{pmatrix}
\Pc_I \\ \Pc_P
\end{pmatrix},\label{eq:suppH}
\end{align}
where $\tilde{\upsilon}=(\upsilon_I+\upsilon_P)/2$ and $\bar{\upsilon}= (\upsilon_I-\upsilon_P)/2$. 
The trace of the matrix, $-\tilde{\delta}$, leads to an average overall decay, $\propto \exp{(-\tilde{\upsilon}t)}$ for each component, which can be taken separately. Then Equation~(\ref{eq:suppH}) can be solved to give
\begin{align}
\nonumber e^{\tilde{\upsilon}t}\Pc_I(t) =& \Pc_I(0) \cos{\Omega t  }
\\ &-\frac{\Pc_I(0)\bar{\upsilon}-\Pc_P(0)\bar{\delta}}{\Omega}\sin{ \Omega t},\label{eq:nonexp1} \\
\nonumber e^{\tilde{\upsilon}t}\Pc_P(t) =& \Pc_P(0) \cos{\Omega t} \\ &-\frac{\Pc_I(0)\bar{\delta}-\Pc_P(0)\bar{\upsilon}}{\Omega}\sin{\Omega t}, 
 \label{eq:nonexp}
\end{align}
with $\Omega=\sqrt{\bar{\delta}^2-\bar{\upsilon}^2}$, and we have assumed $\Omega\neq 0$.

The eigenvalues of the traceless part of the matrix in Equation~(\ref{eq:suppH}) are $\lambda_{\pm} = \pm\sqrt{\bar{\upsilon}^2-\bar{\delta}^2}$, with a degeneracy at $\bar{\delta}=\pm \bar{\upsilon}$. Here we consider the behavior close to the EP at positive $\bar{\delta}=\bar{\upsilon}$. For $\bar{\delta}>\bar{\upsilon}$, both eigenvalues are purely imaginary, and exponentials of the eigenvalues combine to form trigonometric functions of the real $\Omega$. For $\bar{\delta}<\bar{\upsilon}$, $\Omega$ becomes imaginary, and we can replace the trigonometric functions with hyperbolic functions $\cosh{x}=\cos{ix}$, $i\sinh{x}=\sin{ix}$. At the exceptional point $\bar{\delta}^2=\bar{\upsilon}^2$, the solution~(\ref{eq:nonexp1},\ref{eq:nonexp}) is undefined, and the actual solution has non-exponential contributions~\cite{Heiss2010}. Close to this point we can expand Equations~(\ref{eq:nonexp1},\ref{eq:nonexp}) for $\Omega t\ll 1$, with the expansion being accurate for times $t\ll 1/\Omega$. Then as $\Omega\rightarrow 0$, the linear term in the expansion of $\sin{(\Omega t)}/\Omega$ will be independent of $\Omega$, while higher order terms will vanish. The solution,
\begin{eqnarray*}
e^{\tilde{\upsilon}t}\Pc_I(t) = \Pc_I(0) - (\Pc_I(0)-\Pc_P(0))\bar{\upsilon}t + {\cal O}(\Omega), \\
e^{\tilde{\upsilon}t}\Pc_P(t) = \Pc_P(0) - (\Pc_I(0)-\Pc_P(0))\bar{\upsilon}t + {\cal O}(\Omega), 
\end{eqnarray*}
has leading order terms, independent of $\Omega$, which match the exact solution at the exceptional point. Since the amplitudes have leading order terms linear in $t$, $|\Pc|^2$ will be quadratic. Numerically (and as would also be the case experimentally) any small error in the parameters will mean the exceptional point is not precisely reached. However, the exponential nature of the decay will only be apparent at longer times $t\approx 1/\Omega$, as higher order terms become relevant. Since the overall amplitude decays as $e^{-\tilde{\upsilon}t}$, the non-exponential contribution will be dominant as long as $\Omega\ll \tilde{\upsilon}$ ($\approx 0.41\gamma$ for the $20 \times 20$ lattice with $d=0.54\lambda$, which is considered here).

\begin{figure}[htbp]
  \centering
   \includegraphics[width=0.5\columnwidth]{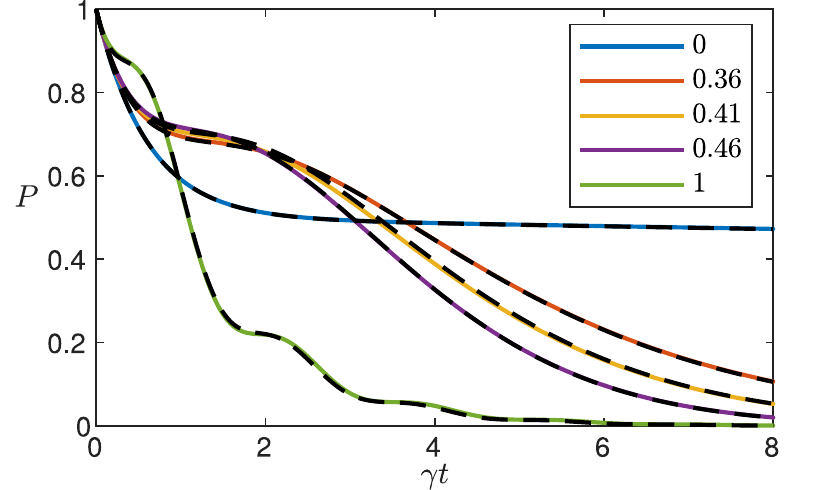}
      \vspace{-0.5cm}
  \caption{Non-exponential decay close to exceptional points. Numerical decay of $P=\sum_{\sigma,j}|\Pc_\sigma^{(j)}|^2$, for initial state $\Pc_I=-\Pc_P=1/\sqrt{2}$, and varying values of $\bar{\delta}/\gamma$ shown in legend. Black dashed lines give respective fits to $\exp(-2\tilde{\upsilon}t)(\cos^2{\Omega t}+\alpha \sin^2{\Omega t})$ predicted by two-mode model. $20\times 20$ lattice with $d=0.54\lambda$.}
  \label{nonexpa}
\end{figure}

In Sec.~\ref{nonexp} we consider a numeric example where the initial condition is $\Pc_I=-\Pc_P=1/\sqrt{2}$, where $\Pc_{I,P}$ are the amplitudes of the collective modes which most closely correspond to the uniform modes in the absence of level shifts. Preparing the lattice in a steady state of this combination, with the level shifts on, and then turning off the drive field, we let this state decay and measure the survival probability $P(t) = \sum_{i,\sigma} |\Pc^{(i)}_\sigma(t)|^2$. The two-mode model predicts that this will be given by the full solution to Equation~(\ref{eq:suppH}), including the overall average decay. We fit the numerical result to the form 
\begin{equation}
P(t) = \exp{(-2\tilde{\upsilon}t)} \left( \cos^2{\Omega t} + \alpha \sin^2{\Omega t}\right),
\end{equation} 
taking $\alpha$ and $\Omega$ to be independent fitting parameters, and only later comparing them to the predicted analytic form. Here $\Omega$ can be either purely real or purely imaginary. The numerical results, along with the numerical best fit, is shown in Figure~\ref{nonexpa}, for various values of $\bar{\delta}$. The extracted numerical parameters for these and additional values of $\bar{\delta}$ are shown in Figure~\ref{fig:decay}. 

\section{Homogeneous equations and Jordan normal form}
\label{appJordan}
For a linear homogeneous system of differential equations, written $\dot{\vec{b}}=i\mathcal{H}\vec{b}$, the solution is given by the matrix exponential
$
\vec{b}(t) = \exp{(i\mathcal{H} t)}\vec{b}(0).
$
If $\mathcal{H}$ is diagonalizable, then $\exp{(i\mathcal{H} t)}$ is diagonal in the same basis, and each eigenvector undergoes independent exponential evolution. At exceptional points where the eigenvectors are not linearly independent however, there is no basis in which $\mathcal{H}$ is diagonal, but it can be written in Jordan normal form, i.e.\ as a block-diagonal matrix $J=V^{-1}\mathcal{H}V$. If an eigenvector $\vec{v}_m$ has geometric multiplicity $M$, there will be a corresponding block $J_m$ consisting of an $M\times M$ matrix with $\lambda_m$ on the diagonal and ones on the superdiagonal~\cite{Franklin1968}. Restricting to the dynamics within one such block, $J_m$, we have in this basis~\cite{Ashida20}
\begin{equation}
\exp{(iJ_m t)} = e^{i\lambda_m t}\left(\id_{M\times M} + \sum_{p=1}^M \frac{(it)^p}{p!} N_{M\times M}^{(p)}	\right),
\end{equation}
where $[\id_{M\times M}]_{ab} =\delta_{ab}$ is the $M\times M$ identity operator and $[N_{M\times M}^{(p)}]_{ab}=\delta_{a,b-p}$ has ones on the $p^\mathrm{th}$ off-diagonal. Then if $M=1$, this reduces to simple exponential behavior, while for $M=2,3,\ldots$ the evolution has terms linear, quadratic, etc., in $t$.
The full dynamics is then easily solved by noting that $\exp{(i\mathcal{H}t)}=V\exp{(iJt)}V^{-1}$ and
\begin{equation}
e^{(iJt)} = \begin{pmatrix}
e^{(iJ_1 t)} && 0 && \ldots && 0 \\
0 && e^{(iJ_2t)} && \ddots && \vdots \\
\vdots && \ddots && \ddots && 0 \\
0 && \ldots && 0 && e^{(iJ_Pt)}
\end{pmatrix},
\end{equation}
where there are a total of $P$ blocks.

\section{One-dimensional scalar electrodynamics}
\label{app1D}

We consider 1D scalar electrodynamics for a planar array when analyzing coherent perfect absorption. This can arise for atoms confined in single-mode waveguides~\cite{Ruostekoski17} or for
uniforms excitations in large planar distributions of atoms~\cite{Facchinetti18,Javanainen19}. For instance, for a uniform distribution of a planar array of atoms we can consider  an average polarization density distributed smoothly in the $yz$ plane
by replacing each dipole $k$ in a uniformly excited lattice $j$, which we write as $\vec{d}_k^{(j)}$, with the average dipole per unit area $\langle \vec{d}_k^{(j)}\rangle/{\cal A}$. This gives the effective polarization
\begin{equation}
P(x) = \frac{1}{{\cal A}}\sum_{j,k}\unitvec{e}_y\cdot \vec{d}_k^{(j)}\delta(x-x_j)=\sum_j \mathfrak{P}^{(j)}\delta(x-x_j),
\end{equation}
which consists of an effective 1D polarization density $\mathfrak{P}^{(j)}$ at each plane $x_j$. The physics is then equivalent to 1D electrodynamics, with $\mathfrak{P}^{(j)}$ corresponding to the polarization amplitude at $x_j$~\cite{Javanainen1999a,Ruostekoski17}. The total field is a sum of the incident field and scattered fields, 
\begin{equation}
\epsilon_0\tilde{E}^{+}(x) = \tilde{D}_F^{+}(x)+\sum_l G(x-x_l)\mathfrak{P}^{(l)},
\end{equation}
with 
\begin{equation}
G(x)=\frac{ik}{2}e^{ik|x|}
\end{equation}
representing the 1D dipole radiation kernel
satisfying $(\nabla^2+k^2)G(x) = -k^2 \delta(x)$.
The equations of motion for the polarization amplitudes then read
\begin{multline}
\dot{\mathfrak{P}}^{(j)} =(i\Delta_j-\gamma_t)\mathfrak{P}^{(j)}+\frac{2i\gamma_{{\rm 1D}}}{k}\tilde{D}_F^{+}(x_j) \\ -\gamma_{{\rm 1D}}\sum_{l\neq j}e^{ik|x_j-x_l|}\mathfrak{P}^{(l)}, 
\end{multline}
where $\Delta_j$ is the detuning of atom $j$,
\begin{equation}
\gamma_{{\rm 1D}}=\frac{k\mathcal{D}^2}{2{\cal A}\hbar\epsilon_0}
\end{equation}
is the radiative 1D linewidth,
and $\gamma_t=\gamma_{{\rm 1D}}+\gamma_l$ includes any additional losses $\gamma_l$, such as loss from the waveguide.
 
\section{${\cal PT}$ symmetry and full reflection}
\label{appref}

\begin{figure}[htbp]
  \centering
   \includegraphics[width=0.5\columnwidth]{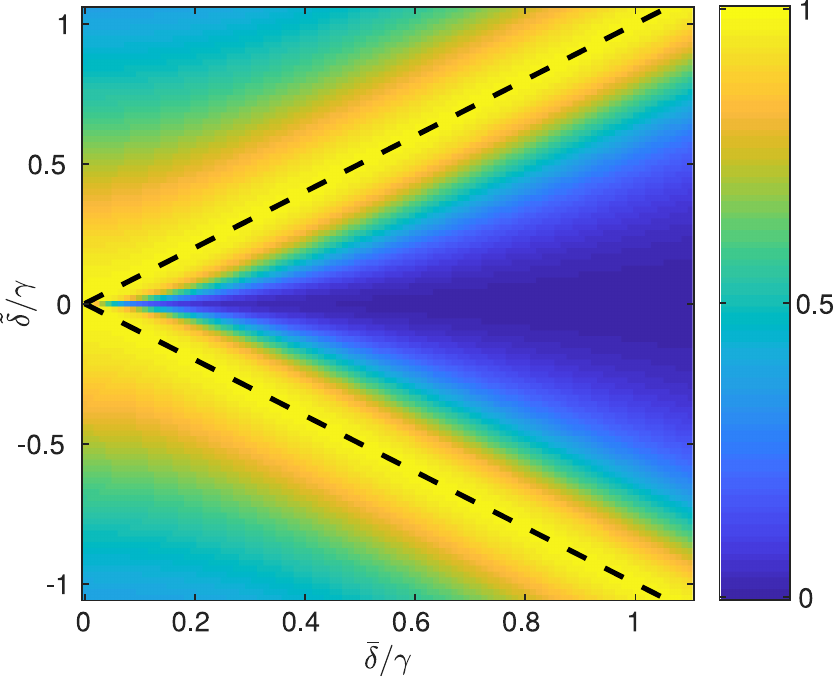}
   \vspace{-0.5cm}
  \caption{Reflectivity $|r|^2$ from a $30\times 30$ lattice with $d=0.54\lambda$. Perfect reflection represented as 
${\cal PT}$ symmetry breaking at the eigenvalues of $H_r$ where $\tilde{\delta}=\pm \bar{\delta}$, illustrated by dashed lines. The exceptional point is at $\bar{\delta}=\tilde{\delta}=0$. }
  \label{h1reflection}
\end{figure}

We show here that also full reflection, i.e.\ where the reflection coefficient $r=-1$ and the transmission coefficient $t=0$, can be represented as 
${\cal PT}$ symmetry breaking. While coherent perfect absorption described in Sec.~\ref{sec:PTtrans} corresponds to a zero eigenvalue of the scattering matrix when the detuning matches a real eigenvalue of $H_T$, full reflection corresponds to an eigenvalue $-1$ in the scattering matrix, and emerges at the eigenvalues of the matrix
\begin{equation}
H_r = \begin{pmatrix}
 i\upsilon_I - i \upsilon_I^c && -i\bar{\delta} \\ i\bar{\delta} &&  i\upsilon_P
\end{pmatrix}.
\end{equation}
This is only observable when the eigenvalue is real, which is possible when $\upsilon_I-\upsilon_I^c=-\upsilon_P$ and $H_r$ is ${\cal PT}$ symmetric. More generally, the condition $\upsilon_I-n\upsilon_I^c=-\upsilon_P$ would lead to an eigenvalue $(n-2)/n$ in the scattering matrix~\cite{Kang2013}.
 
In the large-$N$ limit of strongly confined atoms $\upsilon_P\rightarrow 0$ and $\upsilon_I^c \rightarrow \upsilon_I$.  Then $H_r$ is indeed ${\cal PT}$ symmetric, and has eigenvalues $\tilde{\delta}$ are given by  
\begin{equation}
\det{(H_r-\tilde{\delta}\id)}=\tilde{\delta}^2-\bar{\delta}^2=0.
\end{equation}
In this case the exceptional point is at $\bar{\delta}=0$, and for any value of $\bar{\delta}$ there are two distinct solutions $\tilde{\delta}=\pm\bar{\delta}$ with perfect reflection, which converge at the exceptional point. The calculated reflection from a $30\times 30$ array is shown in Figure~\ref{h1reflection} as a function of $\bar{\delta}$ and $\tilde{\delta}$. Cross-sections of the same reflection are shown in Figure~\ref{fig:twomode}. 
Full reflection can be accurately obtained from the two-mode model,
\begin{equation}\label{eq:ref2mode}
r = \frac{i \upsilon_I Z_P}{\bar{\delta}^2-Z_I Z_P},
\end{equation}
where $Z_{I,P}=(\Delta_0+\delta_{I,P}-\tilde{\delta})+i\upsilon_{I,P}$. For $\upsilon_P\rightarrow 0$ this gives $r=-1$ when $\bar{\delta}^2-\tilde{\delta}^2=0$.

\end{document}